\documentclass[prd,preprint,eqsecnum,nofootinbib,amsmath,amssymb,
               tightenlines,dvips]{revtex4-1}
\usepackage{graphicx}
\input{epsf}
\usepackage{bm}
\usepackage{booktabs}

\def\putbox#1#2{\epsfxsize=#1\textwidth\epsfbox{#2}}
\def\be{\begin{equation}}
\def\ee{\end{equation}}
\def\bea{\begin{eqnarray}}
\def\eea{\end{eqnarray}}
\def\OO{{\cal O}}

\newcommand\Eq[1]{Eq.~(\ref{#1})}

\def\x{{\bm x}}
\def\p{{\bm p}}

\def\k{{\bm k}}
\def\v{{\bm v}}
\def\u{{\bm u}}

\def\naBla{{\bm \nabla}}

\def\la{\label}

\def\slashchar#1{\setbox0=\hbox{$#1$}           
   \dimen0=\wd0                                 
   \setbox1=\hbox{/} \dimen1=\wd1               
   \ifdim\dimen0>\dimen1                        
      \rlap{\hbox to \dimen0{\hfil/\hfil}}      
      #1                                        
   \else                                        
      \rlap{\hbox to \dimen1{\hfil$#1$\hfil}}   
      /                                         
   \fi}                                         %

\def\gsim{\mbox{~{\raisebox{0.4ex}{$>$}}\hspace{-1.1em}
        {\raisebox{-0.6ex}{$\sim$}}~}}
\def\lsim{\mbox{~{\raisebox{0.4ex}{$<$}}\hspace{-1.1em}
        {\raisebox{-0.6ex}{$\sim$}}~}}

\advance\parskip 2pt

\begin {document}


\title
    {
      The Bulk Viscosity of a Pion Gas
    }
\author{Egang Lu and Guy D.\ Moore}
\affiliation
    {%
    Department of Physics,
    McGill University,
    3600 rue University,
    Montr\'{e}al, QC H3A 2T8, Canada
   }%

\date {January 2011}

\begin {abstract}%
    {%
      We compute the bulk viscosity of a gas of pions at temperatures
      below the QCD crossover temperature, for the physical value of
      $m_\pi$, to lowest order in chiral perturbation theory.  Bulk
      viscosity is controlled by number-changing processes which become
      exponentially slow at low temperatures when the pions become
      exponentially dilute, leading to an exponentially large bulk
      viscosity $\zeta \sim (F_0^8/m_\pi^5) \exp(2m_\pi/T)$,
      where $F_0\simeq 93\,$MeV is the pion decay constant.
    }%
\end {abstract}

\maketitle

\section {Introduction}
\la{sec:intro}

One of the most prominent discoveries of the heavy ion program at RHIC
has been the success of hydrodynamics \cite{experiments}
with a zero \cite{ideal_hydro} or very small \cite{nonideal} viscosity.
Though the exact value of the viscosity cannot yet be extracted due to
uncertainties in the initial state and other effects, it
is a robust result that the viscosity near the QCD crossover temperature
is small, $\eta/s < 0.5$ \cite{nonideal}.  On the other hand,
perturbative calculations show that the viscosity to entropy ratio
$\eta/s$ at high temperatures
$T \gg 1$ GeV, where perturbation theory should work, is significantly
higher \cite{AMY6}.  Both theoretical \cite{Complutense} and data-driven
\cite{Venugopalan} analyses of the pion gas indicate that $\eta/s$ also
rises at low temperatures, suggesting that the relative viscosity
bottoms out near the crossover \cite{Csernai}, similar to the behavior
in conventional fluids \cite{KovtunSon}.

The bulk viscosity is also expected to be important in the hydrodynamics
of heavy ion collisions \cite{Heinz}.  Bulk viscosity vanishes for a
conformal system, a good approximation to QCD at high temperatures;
therefore the bulk viscosity to entropy ratio $\zeta/s$ is small at high
temperatures \cite{ArnoldDoganMoore}.  Near the crossover temperature
QCD is very far from conformal, as indicated by the peak in
$(\epsilon-3P)/T^4$ \cite{lattice1,lattice2}, and it is expected that
$\zeta/s$ may display a peak at this scale \cite{Kharzeev,MooreSaremi}.
At lower temperatures QCD is well described by a pion gas.  Existing
studies of pion gases indicate that the bulk viscosity falls away at low
temperatures \cite{Venugopalan,Complutense}.  This suggests that the
ratio $\zeta/s$ shows the opposite behavior of $\eta/s$, peaking near
the transition and falling off to either side
\cite{DIDNTSOMEONESAYTHIS}.

However, previous analyses of the bulk viscosity of a pion gas have been
very incomplete.  In particular, neither standard reference
\cite{Venugopalan,Complutense} considers number changing processes.  But
such processes are essential to the relaxation of particle number to
equilibrium and frequently control the bulk viscosity, as emphasized by
Jeon \cite{Jeon}.  Therefore we believe that what the calculations in
the literature we computing was not really the bulk viscosity of a pion
gas, but the constant for a relaxation process which treated kinetic but
not chemical equilibration.  To make a fair comparison with the
calculations of $\zeta/s$ at higher temperatures one should compute the
true bulk viscosity of a pion gas at low temperatures.  When the bulk
viscosity calculated in this way becomes large, it indicates that the
pion gas will lose chemical equilibrium, a physically interesting
property.

In this paper we will provide a calculation of the bulk viscosity of a
pion gas, including the relaxation via number changing reactions to
chemical equilibrium.  We will work to lowest nontrivial order in chiral
perturbation theory, the effective theory of low energy pions.  That is,
we will make an expansion to lowest order in
$m_\pi / 4\pi F_0$ and $T / 4\pi F_0$, treating $T/m_\pi$ as a free
parameter of order 1.  (Here $F_0$ is the pion decay constant, $m_\pi$
is the pion mass treating the $\pi_0$ and $\pi^{\pm}$ as degenerate, and
$T$ is the temperature as usual.)
Our treatment is therefore only valid at
temperature scales low enough that there are almost no resonances
(such as $\rho$ mesons) and few kaons relative to pions; we will not try
to extrapolate close to the crossover temperature.

In the next section of the paper we will review the physics of bulk
viscosity in a gas of relativistic, massive, weakly coupled bosons,
emphasizing the role played by number changing processes.  We show that
the bulk viscosity is controlled by $m_\pi/T$ and by the rate of number
changing processes.  In Section \ref{sec:chipt} we present the
calculation of the number changing rate within chiral perturbation
theory.  Our numerical results and conclusions are presented in
Section \ref{sec:discussion}, but can be summarized here.  We find that,
as temperature falls, number changing processes become less efficient
and the bulk viscosity actually grows, scaling as
$\zeta/s \sim F_0^{8} T^{-8}$ for $T\sim m_\pi$ and
$\zeta/s \sim F_0^{8} T^{-1/2} m_\pi^{-15/2} \exp(3m_\pi/T)$
for $m_\pi \gg T$.  Therefore the behavior of bulk viscosity is not the
opposite of the behavior of shear viscosity, and in particular both the
bulk viscosity to entropy ratio and the bulk viscosity itself diverge
exponentially in the low temperature limit.

\section{Kinetic description of bulk viscosity}
\label{sec:bulk_kinetic}

By definition, bulk viscosity $\zeta$ is a reduction of the pressure in
an expanding system, and increase in pressure in a contracting system,
proportional to the rate of volume change,%
\footnote{%
    When we write noncovariantly we implicitly work in the instantaneous
    local rest frame.
    We use boldface $\p,\v$ for vectors and normal letters $p,v$ for
    their magnitudes; $P$ is always the pressure.}
\be
\label{def_zeta}
P = P_{\rm eq} - \zeta \naBla \cdot \v
  = P_{\rm eq} - \zeta \frac{dV/dt}{V} \,.
\ee
This arises because the volume change induces a departure from
equilibrium, which in turn modifies the pressure.  To see how this
occurs for a pion gas, we need to describe the system in terms of a
calculable approximation scheme.  Since physical QCD is near the chiral
limit, the pion is a pseudo-Goldstone boson of the (spontaneously but
also explicitly broken) chiral symmetry, and pions are therefore weakly
coupled at low momenta and well described by chiral perturbation theory
(see for instance \cite{GasserLeutwyler,Scherer}).  Weak coupling means that
thermal pions will have well defined quasiparticles which will be well
described by  Boltzmann equations.  Defining the species sum and
integration
\be
\int_{a\p} \equiv \sum_a \int \frac{d^3 \p}{(2\pi)^3 2E_\p} \,,
\ee
the pressure is related to the occupancy of species $a$ at momentum
$\p$, $f_a(\p)$, via
\be
P = \frac{1}{3} \int_{a\p} 2 p^2 f_a(\p) \,.
\label{eq:pressure}
\ee
$f_a(\p)$ in turn evolves according to the Boltzmann equation
\be
\label{eq:Boltzmann}
2E_\p \frac{\partial f_a(\p,t)}{\partial t} + 
2\p \cdot \frac{\partial f_a(\p,t)}{\partial \x}
= - {\cal C}[f] 
= - {\cal C}_{\rm elastic}[f] - {\cal C}_{\rm inel}[f]
\,,
\ee
with ${\cal C}[f]$ the collision operator, which we discuss in more
detail below.

The lefthand side of the Boltzmann equation drives the system from
equilibrium.  Since the bulk viscosity involves one spacetime gradient,
we can find it by expanding the Boltzmann equation to first order in
gradients; since the lefthand side is explicitly first order in
gradients, we may substitute $f_a(\p,t)$ with its equilibrium form
\be
f_0 = \left( \exp\left[\frac{E_\p-\v \cdot \p}{T}\right] - 1 \right)^{-1} \,.
\ee
We take the energy to be
$E = \sqrt{p^2 + m_\pi^2}$, meaning that we will neglect interaction
self-energy corrections in comparison to the explicit pion mass.
Clearly this treatment does not allow us to consider QCD in the strict
chiral symmetry limit, where interaction effects are the only thing
which lead to modified dispersion.  It would be interesting to return to
this case in the future, but we expect it to be rather subtle; for
instance, the lowest order interaction effect actually does not change
the dispersion relation \cite{Gerber,Smilga}, and the next order
only shifts the speed of propagation away from the speed of light
\cite{Gerber,SonStephanov}, which we believe also does not lead
to nonvanishing
bulk viscosity.  Therefore interaction effects appear to arise at a
rather high order in $(T/4\pi F_0)^2$.  Therefore interaction effects
can be neglected for $T \lsim m_\pi$, which is what we are considering.
In this case, explicitly evaluating the lefthand side of the Boltzmann
equation yields
\be
2E_\p \frac{\partial f_a(\p,t)}{\partial t} + 
2\p \cdot \frac{\partial f_a(\p,t)}{\partial \p}
= 2f_0(1{+}f_0) \left( \frac{E^2}{T^2} \frac{dT}{dt}
 + \frac{p_i p_j}{T} \partial_i v_j \right) \,.
\ee
We are interested in the case 
$\partial_i v_j = \frac{1}{3} \delta_{ij} \naBla \cdot \v$.
The temperature changes with time because expansion causes cooling;
at first order in gradients the time dependence of the temperature has
its usual equilibrium relation to $\naBla \cdot \v$,
$dT/dt = -c_s^2 T \naBla \cdot \v$ with
$c_s^2 \equiv dP/d\epsilon$ the squared speed of sound
\cite{ArnoldDoganMoore}.  Therefore the lefthand
side of the Boltzmann equation is
\be
2f_0(1{+}f_0)\; \frac{p^2 - 3 c_s^2 E^2}{3T} \; \naBla \cdot \v \,.
\label{eq:source}
\ee

This ``source'' for departure from equilibrium has no net energy
content.  To see this, note that
\bea
\label{eq:pressure2}
P        = \int_{a\p} \frac{2p^2}{3} f_0(\p) \,,  \qquad
\epsilon & = & \int_{a\p}        2E_\p^2 f_0(\p) \,, \\
\label{eq:cs}
c_s^2 = \frac{dP}{d\epsilon}
= \frac{dP/dT}{d\epsilon/dT} & = &
\frac{ \int_{a\p} \frac{2p^2}{3} \frac{E}{T^2} f_0(1{+}f_0) }
     { \int_{a\p}       2E^2     \frac{E}{T^2} f_0(1{+}f_0) }
\eea
and therefore
\be
\label{speed_sound}
        \int_{a\p} E \: f_0(1{+}f_0) \frac{2p^2}{3T}
= c_s^2 \int_{a\p} E \: f_0(1{+}f_0) \frac{2E^2}{T} \,,
\ee
which shows that there is no energy content for the ``source'' for
departure from equilibrium.  That this occurs is just a check that we
have correctly identified the time dependence of the temperature.
However, the ``source'' {\em does} carry a net particle number,
namely
\be
\left. \frac{dn}{dt} \right|_{\rm LHS} =
\int_{a\p} f_0(1{+}f_0) 2 \frac{p^2 - 3 c_s^2 E^2}{3T} \;
\naBla \cdot \v \neq 0 \,.
\ee
This means that expansion leaves excess pions, relative to the
equilibrium number at the given energy density.  The relaxation of this
excess particle number controls equilibration and bulk viscosity.

Next we turn to the collision term.  At lowest (fourth) order in
$\frac{T,m_\pi}{4\pi F_0}$, the collision term contains only
elastic $\pi\pi \leftrightarrow \pi\pi$ scattering.  Such terms drive
$f_a(\p)$ towards its equilibrium form {\em except} that they cannot
change total particle number.  That is, there is no solution to the
linearized Boltzmann equation with \Eq{eq:source} on the lefthand side
and only $\pi\pi \leftrightarrow \pi\pi$ collision processes on the
righthand side, since the lefthand side includes a change to the net
particle number while the righthand side cannot change particle number.
Therefore a calculation involving only number-conserving processes is
incomplete and inconsistent, as emphasized by Jeon \cite{Jeon} in the
context of scalar $\lambda \phi^4$ theory.%
\footnote{%
    Nevertheless the two previous references on bulk viscosity in a pion
    gas treated only elastic processes.  Ref.~\cite{Venugopalan} got a
    finite answer by using the methodology developed in
    \cite{vanLeeuwen}, which assumes particle number is conserved and
    therefore allows a nonzero chemical potential in the equilibrium
    distribution function.
    Ref.~\cite{Complutense} got a finite answer by doing a one-loop
    diagrammatic evaluation without ``ladder'' graphs, which amounts to
    neglecting the departure from equilibrium in all $f$'s other than
    $f(\p)$ in the Boltzmann equation.}
Therefore we must include as well the lowest order number-changing
process.  Since QCD is parity symmetric but the pion is a parity-odd
scalar, all interaction terms are even in the pion field and the lowest
order kinematically allowed number-changing process is 
$\pi\pi \leftrightarrow \pi\pi\pi\pi$.

At this point there is a simplification.  As in the case of scalar
$\lambda \phi^4$ theory \cite{Jeon} (but unlike the case of weakly
coupled QCD \cite{ArnoldDoganMoore}), number-changing processes are much
less efficient than number-conserving processes in a pion gas.  Number
conserving processes drive the nonequilibrium distribution function
$f(\p) = f_0 + \delta f$ towards an almost-equilibrium form, but with a
chemical potential for particle number,
\be
f^a_\mu(\p) \equiv \left( \exp \frac{E-\mu_a-\p\cdot \v}{T + \delta T} - 1
\right)^{-1} \,.
\ee
Here $\delta T$ is determined by the condition that the energy content
of $f_\mu$ is the same as the energy content of $f_0$.
But number conserving processes
cannot lead to the relaxation of $\mu$ towards zero, because
the elastic collision term vanishes if $f(\p) = f_\mu(\p)$:
\bea
0 = -{\cal C}_{\rm elastic}[f_\mu]
& = & \frac 1{2!} \int_{b\p',c\k d\k'}
  | {\cal M}^{ab,cd}_{\p\p' \rightarrow \k\k'} |^2 
  (2\pi)^4 \delta^4 ( p^\mu+p'{}^\mu-k^\mu-k'{}^\mu ) \qquad \qquad \qquad
\nonumber \\ && \quad \times \;
  \Big( f^a_\mu(\p) f^b_\mu(\p') (1{+}f^c_\mu(\k)) (1{+}f^d_\mu(\k'))
       - \left[ f \leftrightarrow (1{+}f) \right]
\Big)
\qquad
\eea
as the gain term $\propto f(\p)$ and the loss term $\propto (1{+}f(\p))$
cancel.  Therefore $f(\p)$ will equal $f_\mu(\p)$ plus a small
correction.  The value of $\mu$ will dominate the pressure shift.

We cannot make the substitution $f(\p) = f_{\mu}(\p)$ in the elastic
part of the collision operator.  But if we consider the integral
$\int_{a\p}$ of \Eq{eq:Boltzmann}, then the integral
over ${\cal C}_{\rm elastic}$ exactly vanishes, independent of the
form of $f(\p)$.  We {\em can} approximate $f(\p) = f_{\mu}(\p)$ in the
smaller inelastic part of ${\cal C}$, yielding
\be
\naBla \cdot \v \int_{a\p}
          f_0(1{+}f_0) \frac{p^2 - 3 c_s^2 E_\p^2}{3T}
= \int_{a\p} ( -{\cal C}_{\rm inel}[f^a] )
\equiv - C_{\rm inel} \,.
\ee
There are two contributions to this collision term.  One contribution
arises when $\p = \p_1$ is one of the four pions,
\bea
C_{\rm inel}^{4\rightarrow 2}
&=& \frac{1}{3!2!} \int_{a\p_1 b\p_2 c\p_3 d\p_4,e\k_1 f\k_2}
|{\cal M}^{abcd,ef}_{\p_1 \p_2 \p_3 \p_4 \rightarrow \k_1 \k_2} |^2
(2\pi)^4 \delta^4\left(\sum_{i=1..4} p^\mu_i - \sum_{i=1,2} k^\mu_i \right)
\nonumber \\ && \quad \times
\Big( f_\mu^a(\p_1) f_\mu^b(\p_2) f_\mu^c(\p_3) f_\mu^d(\p_4)
       (1{+}f_\mu^e(\k_1)) (1{+}f_\mu^f(\k_2)) - [ f \leftrightarrow
             (1+f) ] \Big) \quad \,.
\label{C_inel}
\eea
The other contribution, $C^{2\rightarrow 4}_{\rm inel}$, arises
when $p=k_1$ is one of the two pions.  It is the
same except $\frac{1}{3!2!}$ is replaced with $-\frac{1}{4!1!}$, so it
cancels half of the above contribution.  (These prefactors
are symmetry factors to eliminate overcounting; for instance, if $b,c,d$
are identical then only $1/3!$ of the phase space should be
integrated over; and if $b,c,d$ are all distinct then the sum
$\sum_{bcd}$ overcounts the possibilities by a factor of $3!$.
The sign difference arises from the relative sign between gain and loss
terms.)

Next we expand $f^a_\mu(\p)$ to first order in $\mu,\delta T$:
\be
\label{first_order_fmu}
f_\mu^a(\p) \simeq f_0(\p) + 
f_0(\p) (1{+}f_0(\p)) \left(\frac{\mu}{T} 
                          + \frac{E \delta T}{T^2} \right) \,.
\ee
Inserting in \Eq{C_inel} and expanding to first order in
$\mu,\delta T$, the distribution functions become
\be
f_0(\p_1) f_0(\p_2) f_0(\p_3) f_0(\p_4) (1{+}f_0(\k_1)) (1{+}f_0(\k_2))
\left( (4-2) \frac{\mu}{T} +
    \left(\sum E_\p - \sum E_k \right) \frac{\delta T}{T^2}
 \right) \,.
\ee
The sum of energies cancels by energy conservation, leaving
\bea
C_{\rm inel}[f_\mu] & = &
\frac{2(2\mu)}{T} \: \frac{1}{4!2!} \int_{a\p_1 b\p_2 c\p_3 d\p_4,
  e\k_1 f\k_2}
|{\cal M}^{abcd,ef}_{\p_1 \p_2 \p_3 \p_4 \rightarrow \k_1 \k_2} |^2
(2\pi)^4 \delta^4\left(\sum_{l=1..4} p^\mu_l - \sum_{l=1,2} k^\mu_l \right)
\nonumber \\ && \quad \times
\Big( f_0(\p_1) f_0(\p_2) f_0(\p_3) f_0(\p_4)
       (1{+}f_0(\k_1)) (1{+}f_0(\k_2)) \Big) \,,
\eea
which determines $\mu$.  The value of $\mu$ in turn determines the
pressure correction,
\be
P - P_{\rm eq} = \int_{a\p} \frac{2p^2}{3} f_0(\p) (1{+}f_0(\p))
\left( \frac{\mu}{T} + \frac{E\delta T}{T^2} \right) \,.
\ee
Recall that $\delta T$ is set by the condition that the perturbation
carry no net energy, which using \Eq{first_order_fmu} is
\be
\delta T \int_{a\p} f_0(1{+}f_0) \frac{2E^3}{T^2}
 = -\mu  \int_{a\p} f_0(1{+}f_0) \frac{2E^2}{T} \,.
\ee
Together with \Eq{speed_sound} means
\be
P - P_{\rm eq} = \mu \int_{a\p} f_0(1{+}f_0) 2 \frac{p^2 - 3c_s^2 E^2}{3T} \,.
\ee
Putting everything together with the definition \Eq{def_zeta}, we find
\bea
\label{eq:zeta}
\zeta & = & \frac{ T \left(
            \int_{a\p} f_0(1{+}f_0) 2 \frac{p^2 - 3c_s^2 E^2}{3T}
                   \right)^2 }
            { 4 \hat{C}_{\rm inel}} \,,
 \\
\hat{C}_{\rm inel} & = & 
    \frac{1}{4!2!} \int_{a\p_1 b\p_2 c\p_3 d\p_4, e\k_1 f\k_2}
  |{\cal M}^{abcd,ef}_{\p_1 \p_2 \p_3 \p_4 \rightarrow \k_1 \k_2} |^2
(2\pi)^4 \delta^4 \left( \sum_{i=1..4} p^\mu_i - \sum_{i=1,2} k^\mu_i \right)
\nonumber \\ && \qquad \times
\Big( f_0(\p_1) f_0(\p_2) f_0(\p_3) f_0(\p_4)
       (1{+}f_0(\k_1)) (1{+}f_0(\k_2)) \Big) \,.
\label{Cinel}
\eea
The integration in the numerator is elementary, so evaluating the
denominator will be our main challenge.

Using the technique developed in \cite{AMY1,AMY6,ArnoldDoganMoore},
we would arrive at the same result by using the single parameter
{\it Ansatz} for the departure from equilibrium shown in
\Eq{first_order_fmu}.  In the notation used there, each term in the
numerator is $\tilde{S}$ and the denominator is $\tilde{C}$.  The
factor of 4 is essentially $(\mu+\mu+\mu+\mu-\mu-\mu)^2/\mu^2$ and can be
understood as follows; each number changing collision changes
particle number by 2 (one factor of 2), and a chemical potential
makes the forward process faster than the backwards process by
$2\mu/T$ (the other factor of 2).

\section{Chiral perturbation theory}
\label{sec:chipt}

Quantum Chromodynamics is considered as the fundamental theory for
describing strong interactions between quarks and gluons. However, at
energies below the breaking scale of chiral symmetry, quarks and
gluons are confined within the asymptotic hadron states, such as pions,
kaons, and $\eta$ mesons. In this energy regime, the QCD coupling constant
becomes so large that the theory is highly non-perturpative and we still
lack an analytical method to solve it.
However, the situation gets better if we write an effective field theory
describing the meson states.  It is an experimental fact that, at
sufficiently low energies, the light mesons interact weakly with each
other, with the strength of interactions controlled by a derivative
expansion which is described by Chiral Perturbation Theory
\cite{Weinberg,Scherer}, an effective theory for the interactions of
light pseudoscalar mesons.

In the chiral limit, the QCD Lagrangian possesses a $SU\left( N\right)_L
\times SU \left(N\right)_R\times U\left( 1 \right)_V$ global
symmetry. Here $N$ denotes the number of flavors. The axial symmetry 
$U \left( 1 \right)_A$ of the QCD Lagrangian, present at the classical
level, is broken due to a quantum anomaly. Experimental facts, such as
the hadron spectrum and quark condensate, indicate that the 
$SU \left( N\right)_L \times SU \left(N\right)_R 
 \times U \left( 1 \right)_V$ spontaneously breaks down
into $SU \left( N \right)_V\times U \left( 1 \right)_V$. According to
Goldstone's Theorem, in this process, massless Goldstone bosons, which
are identified with the pseudoscalar mesons, are generated. Since we are
dealing with a pure pion gas, we only focus on the case that $N=2$, that
is, only up and down quarks are of concern in our discussion.

In this specific case, the three kinds of pions are considered as the
Goldstone bosons, and they transform as a triplet under the subgroup $SU_V
\left( 2 \right)$. Moreover, pion fields, the three-component vector $\vec{\Phi} = \left( \phi_1, \phi_2,
  \phi_3 \right)$, are isomorphic to the quotient group $SU \left( 2
\right)_L\times SU \left( 2 \right)_R/SU \left( 2 \right)_V$.

In the chiral limit, one can, in terms of pion fields 
$\vec{\Phi} = \left( \phi_1, \phi_2, \phi_3 \right)$, construct the general
Langrangian invariant under $SU \left( 2 \right)_L\times SU \left( 2
\right)_R\times U \left( 1 \right)_V$, with the ground state invariant
only under subgroup $SU \left( 2 \right)_V\times U \left( 1 \right)
_V$. But in fact, instead of being massless, pions have small but finite
masses around 135 MeV. This is because chiral symmetry is not an exact
one. It is broken by a small amount due to the nonvanishing masses of up and down
quarks. In order to give masses to pions, one also needs to add an explicit
symmetry breaking term into the Lagrangian, which is treated as a small
perturbation.

The general effective Lagrangian can be organized by chiral order,
\begin{displaymath}
\mathcal{L}_\text{eff} =\mathcal{L}_2 +\mathcal{L}_4 +\mathcal{L}_6 +\cdots
\end{displaymath}
where the subscripts indicate the chiral order. $\mathcal{L}_2$ with the
smallest chiral order contains the minimum number of derivatives and
quark mass terms.  It reads \cite{Scherer}
\begin{equation}
\mathcal{L}_2 =-\frac{F^2_0}{4}\text{Tr} 
         \left(  \partial_\mu U \partial^\mu U^\dagger \right)
      +\frac{F^2_0 m^2_\pi}{4} \text{Tr} \left( U+U^\dagger \right) \, .
\end{equation}
Here
\begin{eqnarray}
U & = & \exp \left( i \frac{\vec{\tau} \cdot \vec{\Phi}}{F_0} \right) 
    =\exp \left[ i \frac{\phi \left( x \right)}{F_0} \right] \, , \\
\phi \left( x \right) & = & \left( \begin{array}{cc} \phi_3 
             & \phi_1 - i \phi_2 \\ \phi_1 + i \phi_2 &
             - \phi_3 \end{array} \right) 
    = \left ( \begin{array}{cc} \pi^0 & \sqrt{2} \pi^+ 
       \\ \sqrt{2} \pi^- & - \pi^0 \end{array} \right) \,,
\end{eqnarray}
where $F_0\approx 93\text{MeV}$ is the pion decay constant and
$\vec{\tau}$ are the three Pauli matrices.

\begin{figure}
\hfill
\putbox{0.63}{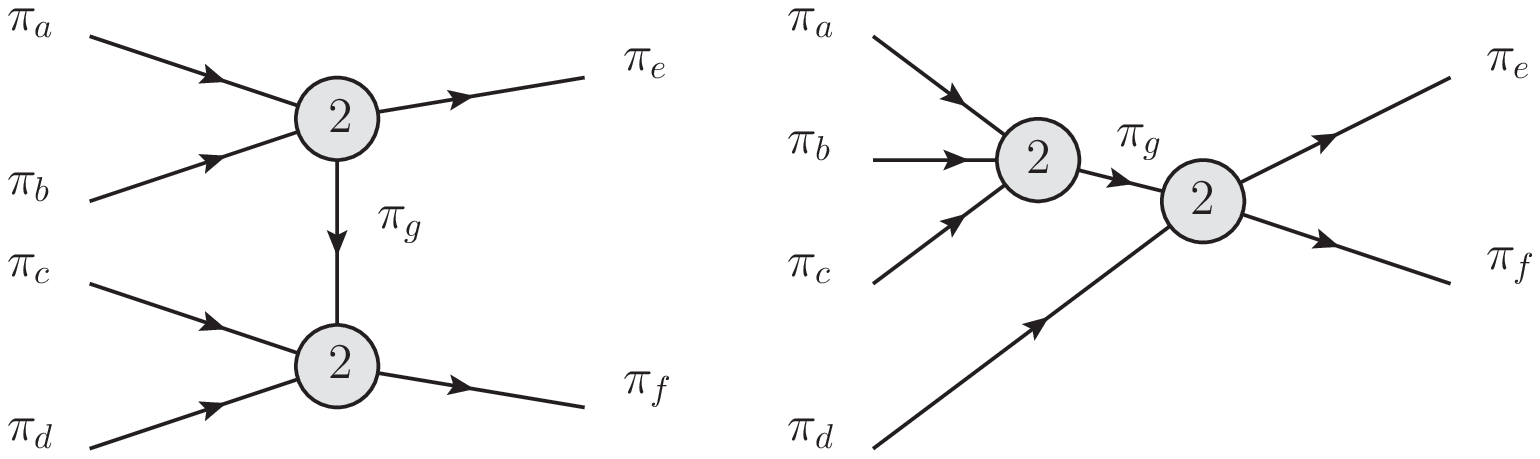}
\hfill
\hfill
\hfill
\putbox{0.32}{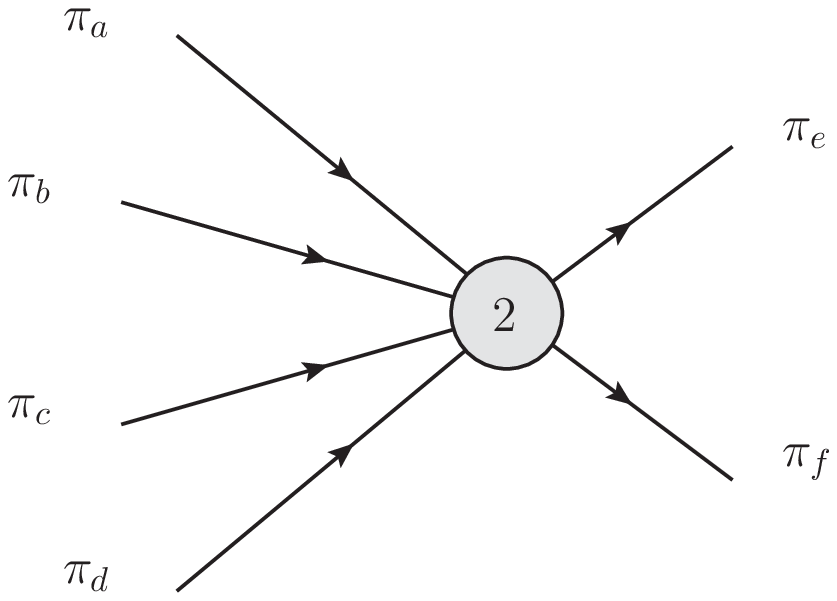}
\hfill $\vphantom{.}$
\caption[Diagrams]
{\label{fig:diagrams}
 Three classes of diagrams needed to evaluate the inelastic scattering
 rate to lowest order in chiral perturbation theory.  Here the Roman 
 subscripts are the Cartesian isospin indices and the number
 ``2'' in a circle denotes the chiral order of the vertex.}
\end{figure}

The matrix element for elastic scattering is well known in chiral
perturbation theory \cite{Scherer} and does not concern us, since
\Eq{eq:zeta} shows that the bulk viscosity is controlled by
number-changing processes.  
We need the matrix element for $4\pi \rightarrow 2\pi$ processes.
Three classes of diagrams can arise, as depicted in Figure
\ref{fig:diagrams}.  For each class, we must sum over the distinct
permutations of the external lines.

Expanding $\mathcal{L}_2$, one can find the corresponding matrix
elements.  For the representative permutations shown in
Figure \ref{fig:diagrams}, the matrix elements read (here for simplicity of 
writing down the matrix elements, all the four-momenta are viewed as incoming)
\begin{eqnarray}
\mathcal{M}_1 & = &  \sum\limits_{g=1,2,3} 
     \mathcal{V}\left( a,b,e,g \right) 
     \frac{-i}{p_g^2 + m_\pi^2} \mathcal{V} \left( c,d,f,g \right)\\
\mathcal{M}_2 & = & \sum\limits_{g=1,2,3} 
      \mathcal{V} \left( a,b,c,g \right) 
      \frac{-i}{p_g^2 + m_\pi^2} \mathcal{V} \left( d,e,f,g \right)\\
\mathcal{M}_3 & = & \frac{i}{9F_0^4} \delta^{ab} \delta^{cd} 
        \delta^{ef} \left[ 4 \left( p_a \cdot p_b 
       + p_c \cdot p_d + p_e \cdot p_f \right) 
       - 3m_\pi^2 \right] \nonumber \\ 
&&\quad {}+ \mbox{all distinct pairings of the set }\{a, b, c, d, e, f\}\,,
\end{eqnarray} 
where $\sum\limits_{g=1,2,3}$ is a sum over the species type in the propagator,
$p_g$ is the four-momentum of the propagator and
\begin{eqnarray}
\mathcal{V} \left( \alpha ,\beta ,\gamma ,g \right)
&=&
         \left( i/3F_0^2 \right) 
       \Big[ \phantom{+} \delta^{\alpha g} \delta^{\beta \gamma} 
        \left( 2p_\alpha \cdot p_\beta 
             + 2 p_\alpha \cdot p_\gamma -4p_\beta \cdot p_\gamma + m_\pi^2\right) \nonumber 
\\ & & \phantom{(i/3F_0^2) \Big[} + 
    \delta^{\beta g} \delta^{\alpha \gamma}  
       \left( 2p_\alpha \cdot p_\beta 
          + 2 p_\beta \cdot p_\gamma - 4p_\alpha \cdot p_\gamma + m_\pi^2\right)
\nonumber \\ & & \phantom{(i/3F_0^2) \Big[} + 
    \delta^{\gamma g} \delta^{\alpha \beta} 
       \left( 2p_\alpha \cdot p_\gamma 
     + 2 p_\beta \cdot p_\gamma - 4p_\alpha \cdot p_\beta + m_\pi^2\right) \Big] \,.
\end{eqnarray}
Therefore, the transition amplitude of the lowest order in question is
\begin{equation}
|\mathcal{M}|^2 = \left| \sum\limits_{\rm perm} \mathcal{M}_1 
    +\sum\limits_{\rm perm} \mathcal{M}_2 + \mathcal{M}_3 \right|^2 \, ,
\end{equation}
where $\sum\limits_{\rm perm}$ means a sum is taken over all distinct
permutations of the external lines.

This transition amplitude has a very complicated form, so we cannot
finish the integral $\hat{C}_{\rm inel}$ analytically. Therefore, we resort to
numerical methods. For the numerical calculation,
we work in the local plasma rest frame, that is, the rest-frame four velocity
is $u^\mu = \left( 1,0,0,0 \right)$.
The distribution function in this frame is just $f_0 \left( \vec{p} \right)=
\left[ \exp \left( E_p \right) -1 \right]^{-1}$. The main challenge is
to perform the phase space integration over 6 external states.
We consider the process as $4 \pi \rightarrow 2 \pi$, that is
four incoming paticles and two outgoing particles.  We perform
unconstrained integrations over the four incoming particle momenta in
spherical coordinates with $\p_a$ as the $z$ axis and $\p_b$ lying in
the $x,z$ plane,
\begin{equation}
\int \frac{d^3 \bm{p}_a d^3 \bm{p}_b d^3 \bm{p}_d d^3 \bm{p}_e}
          {\left( 2 \pi \right)^{12} 16 E_a E_b E_c E_d}  =
\frac{1}{8 \left( 2 \pi \right)^{10}}
\int \frac{p_a^2 d p_a}{E_a} 
\frac{p_b^2 d p_b d \theta_{b}}{E_b} 
\frac{p_b^2 d p_b d \Omega_{c}}{E_{c}} 
\frac{p_b^2 d p_b d \Omega_{d}}{E_d}
\end{equation}
and then apply the energy-momentum conserving delta function to simplify
the two-particle final phase space integration in the manner
shown in \cite{Byckling,Kersevan}. The final state
phase space can be rewitten as
\begin{equation}
\int \frac{d^3 \bm{p}_e d^3 \bm{p}_f}{\left( 2 \pi \right)^6 4 E_e E_f} 
\delta^4 \left( p_a + p_b + p_c + p_d - p_e - p_f \right) 
  = \frac{\sqrt{1-4m^2/s}}{2^9 \pi^6}
\int d \Omega^*
\end{equation}
where $\Omega^*$ is defined in the center of mass frame of the total
incoming momentum
$k^\mu=p_a^\mu+p_b^\mu+p_c^\mu+p_d^\mu$, and $s=-k^2$ is the
Mandelstam variable. In the center of mass frame it is most convenient
to work in spherical coordinates with the $z$ axis chosen along the
boost axis to the plasma rest frame.
All dot products between incoming momenta are easily expressed in terms
of the plasma frame variables, as is the Mandelstam variable $s$.
For final state particle energies and dot products between an incoming
and an outgoing momentum, we need to apply the boost between center of
mass and plasma frame variables.
An alternative approach is to consider the process $2 \pi \rightarrow 4
\pi$ and apply the energy-momentum conserving delta function on the
4-particle final state phase space as shown in
\cite{Byckling,Kersevan}; but this approach is a little
more involved.  The resulting 11-dimensional integrations are performed
by Monte-Carlo integration using CUBA \cite{CUBA}.

We determine the pressure, speed of sound, and numerator of \Eq{eq:zeta}
by performing the integrals in \Eq{eq:pressure2}, \Eq{eq:cs},
and \Eq{eq:zeta} numerically.
It has become customary to compare viscosities with
the entropy density $s = \partial P / \partial T$, which has the same
units as $\zeta$.  Differentiating \Eq{eq:pressure2},
\begin{equation}
\label{eq:entropy}
s = \int_{a\p} \frac{2Ep^2}{3T^2} f_0(1{+}f_0)
\end{equation}
which we also handle numerically.

\section{Results and discussion}
\label{sec:discussion}

\begin{figure}
\hfill
\putbox{0.46}{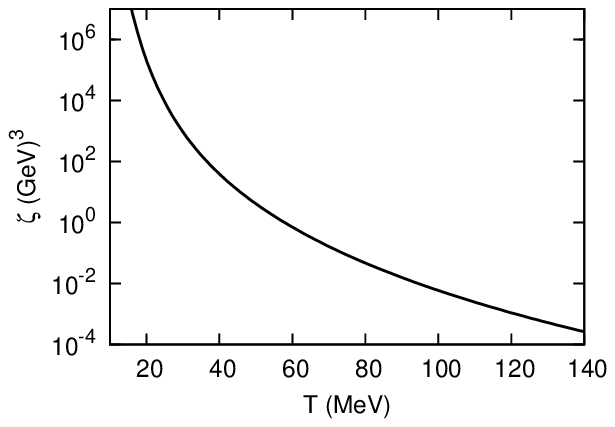}
\hfill
\putbox{0.46}{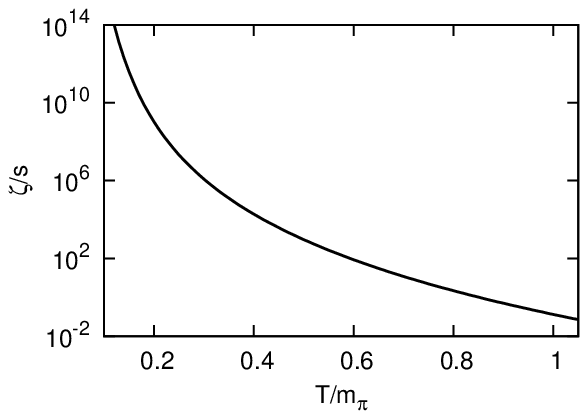}
\hfill $\vphantom{.}$
\caption{The numerical calculation of bulk viscosity $\zeta$ and the
bulk viscosity to entropy ratio $\zeta/s$.}
{\label{fig:numerical}}
\end{figure}

The results of numerical calculation of the bulk viscosity are shown in
the Table \ref{tab:table} and Figure \ref{fig:numerical}.
The most obvious feature of the bulk viscosity is that
$\zeta$ and $\zeta/s$ both rise as the
temperature is lowered.  This is the same behavior as the shear
viscosity, in contrast to the high temperature regime,
$T \gg \Lambda_{\rm QCD}$, where $\eta/s$ rises but $\zeta/s$ falls with
rising temperature.

\begin{table}

\begin{tabular}{|cccccccc|}

\hline
$T \,\left( \text{MeV} \right)$ & 10 & 20 & 30 & 40 & 50 & 60 & 70  \\
\hline
$\zeta\, \left( \text{GeV}^3 \right) $ & $\,3.6\times 10^{11}\,$ & $\,2.1\times 10^{5}\,$ & $\,9.3\times 10^{2}\,$  & $\,3.9 \times 10^{1}\,$ & $\,4.1\times 10^0\,$ & $\,7.0\times 10^{-1}\,$ & $\,1.6\times 10^{-1}\,$ \\
\hline
$s \, \left( \text{GeV}^3 \right)$ & $\,2.4\times 10^{-10}\,$ & $\,3.9\times 10^{-7}\,$ & $\,5.8 \times 10^{-6}\,$ & $\,2.7 \times 10^{-5}\,$ & $\,7.4 \times 10^{-5}\,$ & $\,1.6 \times 10^{-4}\,$ & $2.9\times 10^{-4}$\\
\hline
\hline
$T \,\left( \text{MeV} \right)$ & 80 & 90 & 100 & 110 & 120 & 130 & 140  \\
\hline
$\zeta\, \left( \text{GeV}^3 \right) $ & $\,4.7\times 10^{-2}\,$ & $\,1.6\times 10^{-2}\,$ & $\,5.9\times 10^{-3}\,$  & $\,2.4 \times 10^{-3}\,$ & $\,1.1\times 10^{-3}\,$ & $\,5.2\times 10^{-4}\,$ & $\,2.6\times 10^{-4}\,$ \\
\hline
$s \, \left( \text{GeV}^3 \right)$ & $\,4.7\times 10^{-4}\,$ & $\,7.1\times 10^{-4}\,$ & $\,1.0 \times 10^{-3}\,$ & $\,1.4 \times 10^{-3}\,$ & $\,1.9 \times 10^{-3}\,$ & $\,2.5 \times 10^{-3}\,$ & $3.2\times 10^{-3}$\\
\hline
\end{tabular}
\caption{\label{tab:table}Values of $\zeta$ and $s$ at certain temperatures}
\end{table}

We can understand the rising behavior of $\zeta/s$ with lower temperature,
for $T \sim m_\pi$, as follows.  First, the strength of conformal symmetry
breaking depends on $m_\pi/T$, which gets larger as $T$ gets smaller.
Second, as the temperature gets lower, the typical momentum scale for
pions gets lower.  Since pions are pseudoGoldstone bosons, they interact
mostly through high-derivative interactions, which get weaker as the
energy scale is lowered.  Therefore the system remains out of
equilibrium longer, leading to higher viscosities.  This last effect
becomes very important when $T \ll m_\pi$.  In this case the density of
pions falls exponentially, $n \sim (m_\pi T)^{3/2} \exp(-m_\pi/T)$.  The
probability to have {\sl four} pions in one place at one time, to
participate in a number-changing collision, is therefore {\sl
  exponentially} small,%
\footnote{%
    Or for the inverse process, the probability to have two pions
    with enough energy to generate four pions is exponentially small}
so the rate of number changing processes is exponentially suppressed and
the bulk viscosity becomes exponentially large.  This behavior was
pointed out in the context of scalar field theory by Jeon \cite{Jeon}.

In the low temperature limit $T\ll m_\pi$, the
behavior of the bulk viscosity can be calculated analytically. In this
regime the distribution function for incoming pions is well approximated
by the nonrelativistic form 
$f_0(p) \simeq e^{-m/T} e^{-p^2/2m_\pi T}$.  The
typical value of the momentum $p$ is $p \sim \sqrt{m_\pi T} \ll m_\pi$,
which greatly simplifies both the initial particle phase space and the
matrix element.  For the
purposes of evaluating the matrix element ${\cal M}_{4\rightarrow 2}$,
at leading order we can make the approximation that
\begin{equation}
p_a=p_b=p_c=p_d=\left( m,\vec{0} \right) , \quad
p_e=\left( 2m, \sqrt{3} \vec{m} \right) \quad \text{and} 
    \quad p_f=\left( 2m, -\sqrt{3} \vec{m} \right) \,.
\end{equation}
Under this approximation the summation of matrix element over species can be found in closed form:
$\sum\limits_{a,b,c,d,e,f} \left| \mathcal{M} \right|^2 \sim 2025m_\pi^4/2F_0^8$.  Factoring it
out of the integral, and approximating $s \sim 16m^2$, the remaining
angular integrations can be performed easily.
Then the phase space integral in $\hat{C}_{\rm inel}$ reduces to
\begin{eqnarray}
&& \int \frac{d^3 \bm{p}_a d^3 \bm{p}_b d^3 \bm{p}_c d^3 \bm{p}_d d^3 \bm{p}_e d^3 \bm{p}_f}
   {\left( 2 \pi \right)^{18} 64 E_a E_b E_c E_d E_e E_f} 
   \left( 2 \pi \right)^4 \delta^4 \left( p_a +p_b+p_c
                                           +p_d-p_e-p_f \right)
\nonumber \\ && \hspace{2in} {} \times
   f_0(p_a)f_0(p_b) f_0(p_c) f_0(p_d) (1{+}f_0(p_e))(1{+}f_0(p_f))
   \hspace{0.3in}
\\ & \simeq &  \frac{\sqrt{3}}{4096 \pi^9 m_\pi^4 } 
      \int p_a^2 dp_a p_b^2 dp_b p_c^2 dp_c p_d^2 dp_d
    \; e^{-4m_\pi/T} e^{-(p_a^2+p_b^2+p_c^2+p_d^2)/2m_\pi T}
\nonumber \\
&=& \frac{\sqrt{3} m_\pi^2 T^6 e^{-4m_\pi/T}}{16384 \pi^{7}}
 \,.
\end{eqnarray}
We also need to carry out the integral in \Eq{eq:zeta}, which includes
determining the speed of sound from \Eq{speed_sound}.  Here there is a
subtlety; if we compute $c_s^2$ to lowest order and put it in
\Eq{eq:zeta}, again computing in the nonrelativistic approximation, we
get zero.  Both equations must be expanded to second order in $T/m_\pi$,
yielding
\begin{eqnarray}
c_s^2 & = & \frac{T}{m_\pi} - \frac{T^2}{2m_\pi^2} + \OO(T^3/m_\pi^3) \,,
\\
\int_{a\p} f_0(1{+}f_0) 2\frac{p^2-3c_s^2 E^2}{3T}
  & = &
  \exp(-m_\pi/T) \times \left( -3 \frac{m_\pi^{1/2} T^{5/2}}{(2\pi)^{3/2}} 
                               + \OO(T^{7/2} m_\pi^{-1/2}) \right) \,,
\end{eqnarray}
where the factor of 3 counts the number of pion species.
Combining these results, the low temperature limit of the bulk viscosity
is
\begin{equation}
\zeta(T\ll m) \simeq \frac{16384 \sqrt{3} \pi^4}{225} 
   \frac{F_0^8}{m_\pi^5} \exp \frac{2m_\pi}{T} \, , \qquad
\frac{\zeta}{s}(T\ll m) \simeq \frac{32768\sqrt{6}
              \pi^{\frac{11}{2}} F_0^8}
        {675 m_\pi^{\frac{15}{2}} T^{\frac 12}} 
        \exp \frac{3m_\pi}{T} \,,
\end{equation}
where we used the leading order behavior of \Eq{eq:entropy},
$s \simeq (3m_\pi^{\frac 52}T^{\frac 12}/(2\pi)^{\frac 32}) \exp(-m_\pi/T)$.
These low temperature asymptotics are consistent with our numerical
results.

We should emphasize that at temperatures such that the bulk viscosity is
very large, $\zeta \naBla \cdot \u \gsim P$, the near-equilibrium
expansion implicit in defining and using $\zeta$ has broken down.  When
this occurs, the system in question has fallen out of chemical
equilibrium; in fact $\zeta \naBla \cdot \u > P$ can be taken as a
criterion for the breakdown of chemical equilibrium and the freezing out
of number changing processes.  And when $\zeta$ becomes 
{\sl exponentially} large, the approximation that we treat QCD without
including electromagnetic interactions ceases to be valid.  At low
temperatures the dominant number changing process would actually be
$\pi^0 \rightarrow 2\gamma$ (and its crossings).  We will not
consider this extension here.

In conclusion, we have computed the bulk viscosity of a pion gas, the
natural low-temperature limit of QCD.  We find that the bulk viscosity
rises at low temperatures, growing exponentially as
$\zeta \sim \exp(2m_\pi/T)$ in the $T\ll m_\pi$ limit.  This growth
implies that kinetic theory will generally break down at low
temperatures, explaining chemical freezeout.

\section*{Acknowledgements}

We would like to thank Aleksi Kurkela and Yi Wang for useful
conversations. This work was supported in part by the Natural Sciences
and Engineering Research Council of Canada.


\begin {thebibliography}{39}

\bibitem{experiments}
  K.~Adcox {\it et al.}  [PHENIX Collaboration],
  Nucl.\ Phys.\  A {\bf 757} (2005) 184;
  B.~B.~Back {\it et al.} [PHOBOS Collaboration],
  Nucl.\ Phys.\  A {\bf 757} (2005) 28;
  I.~Arsene {\it et al.}  [BRAHMS Collaboration],
  Nucl.\ Phys.\  A {\bf 757} (2005) 1;
  J.~Adams {\it et al.}  [STAR Collaboration],
  Nucl.\ Phys.\  A {\bf 757} (2005) 102.

\bibitem{ideal_hydro}
 D.~Teaney, J.~Lauret and E.~V.~Shuryak,
  Phys.\ Rev.\ Lett.\  {\bf 86} (2001) 4783;
  P.~Huovinen, P.~F.~Kolb, U.~W.~Heinz, P.~V.~Ruuskanen and S.~A.~Voloshin,
  Phys.\ Lett.\  B {\bf 503} (2001) 58;
  P.~F.~Kolb, U.~W.~Heinz, P.~Huovinen, K.~J.~Eskola and K.~Tuominen,
  Nucl.\ Phys.\  A {\bf 696} (2001) 197;
  T.~Hirano and K.~Tsuda,
  Phys.\ Rev.\  C {\bf 66} (2002) 054905;
  P.~F.~Kolb and R.~Rapp,
  Phys.\ Rev.\  C {\bf 67} (2003) 044903.

\bibitem{nonideal}
P.~Romatschke and U.~Romatschke,
  Phys.\ Rev.\ Lett.\  {\bf 99}, 172301 (2007)
  [arXiv:0706.1522 [nucl-th]];
M.~Luzum and P.~Romatschke,
  Phys.\ Rev.\  C {\bf 78}, 034915 (2008)
  [Erratum-ibid.\  C {\bf 79}, 039903 (2009)]
  [arXiv:0804.4015 [nucl-th]];
K.~Dusling and D.~Teaney,
  Phys.\ Rev.\  C {\bf 77}, 034905 (2008)
  [arXiv:0710.5932 [nucl-th]];
H.~Song and U.~W.~Heinz,
  Phys.\ Rev.\  C {\bf 77}, 064901 (2008)
  [arXiv:0712.3715 [nucl-th]].

\bibitem{AMY6}
P.~B.~Arnold, G.~D.~Moore and L.~G.~Yaffe,
  JHEP {\bf 0305}, 051 (2003)
  [arXiv:hep-ph/0302165].


\bibitem{Complutense}
D.~Fernandez-Fraile and A.~G.~Nicola,
  Phys.\ Rev.\ Lett.\  {\bf 102}, 121601 (2009)
  [arXiv:0809.4663 [hep-ph]].

\bibitem{Venugopalan}
M.~Prakash, M.~Prakash, R.~Venugopalan and G.~Welke,
  Phys.\ Rept.\  {\bf 227}, 321 (1993).

\bibitem{Csernai}
L.~P.~Csernai, J.~I.~Kapusta and L.~D.~McLerran,
  Phys.\ Rev.\ Lett.\  {\bf 97}, 152303 (2006)
  [arXiv:nucl-th/0604032].

\bibitem{KovtunSon}
P.~Kovtun, D.~T.~Son and A.~O.~Starinets,
  JHEP {\bf 0310}, 064 (2003)
  [arXiv:hep-th/0309213].


\bibitem{Heinz}

 H.~Song and U.~W.~Heinz,
  Nucl.\ Phys.\  A {\bf 830}, 467C (2009)
  [arXiv:0907.2262 [nucl-th]];
  Phys.\ Rev.\  C {\bf 81}, 024905 (2010)
  [arXiv:0909.1549 [nucl-th]].
 K.~Rajagopal and N.~Tripuraneni,
  JHEP {\bf 1003}, 018 (2010)
  [arXiv:0908.1785 [hep-ph]].


\bibitem{ArnoldDoganMoore}
  P.~B.~Arnold, C.~Dogan and G.~D.~Moore,
  Phys.\ Rev.\  D {\bf 74}, 085021 (2006)
  [arXiv:hep-ph/0608012].

\bibitem{lattice1}
 A.~Bazavov {\it et al.},
  Phys.\ Rev.\  D {\bf 80}, 014504 (2009)
  [arXiv:0903.4379 [hep-lat]];
%
 M.~Cheng {\it et al.},
  Phys.\ Rev.\  D {\bf 81}, 054504 (2010)
  [arXiv:0911.2215 [hep-lat]].

\bibitem{lattice2}

S S.~Borsanyi {\it et al.},
  JHEP {\bf 1011}, 077 (2010)
  [arXiv:1007.2580 [hep-lat]].

\bibitem{Kharzeev}
 D.~Kharzeev and K.~Tuchin,
  JHEP {\bf 0809}, 093 (2008)
  [arXiv:0705.4280 [hep-ph]];
 F.~Karsch, D.~Kharzeev and K.~Tuchin,
  Phys.\ Lett.\  B {\bf 663}, 217 (2008)
  [arXiv:0711.0914 [hep-ph]].

\bibitem{MooreSaremi}
  G.~D.~Moore and O.~Saremi,
  JHEP {\bf 0809}, 015 (2008)
  [arXiv:0805.4201 [hep-ph]].

\bibitem{DIDNTSOMEONESAYTHIS}
J.~W.~Chen and J.~Wang,
  Phys.\ Rev.\  C {\bf 79}, 044913 (2009)
  [arXiv:0711.4824 [hep-ph]].


\bibitem{Jeon}
  S.~Jeon,
  Phys.\ Rev.\  D {\bf 52}, 3591 (1995)
  [arXiv:hep-ph/9409250].

\bibitem{GasserLeutwyler}
J.~Gasser and H.~Leutwyler,
  Annals Phys.\  {\bf 158}, 142 (1984).

\bibitem{Scherer}
  S.~Scherer,
  Adv.\ Nucl.\ Phys.\  {\bf 27}, 277 (2003)
  [arXiv:hep-ph/0210398].

\bibitem{Gerber}
  P.~Gerber and H.~Leutwyler,
  Nucl.\ Phys.\  B {\bf 321}, 387 (1989).

\bibitem{Smilga}
A.~V.~Smilga,
  Phys.\ Rept.\  {\bf 291}, 1 (1997)
  [arXiv:hep-ph/9612347].

\bibitem{SonStephanov}
D.~T.~Son and M.~A.~Stephanov,
  Phys.\ Rev.\ Lett.\  {\bf 88}, 202302 (2002)
  [arXiv:hep-ph/0111100].
  Phys.\ Rev.\  D {\bf 66}, 076011 (2002)
  [arXiv:hep-ph/0204226].

\bibitem{vanLeeuwen}
W.~A.~Van Leeuwen, P.~H.~Polack and S.~R.~de Groot, Physica 
{\bf 63}, 65 (1973).

\bibitem{AMY1}
  P.~B.~Arnold, G.~D.~Moore and L.~G.~Yaffe,
  JHEP {\bf 0011}, 001 (2000)
  [arXiv:hep-ph/0010177].
\bibitem{Weinberg}
Weinberg, The Quantum Theory of Fields vol2, Cambridge, 1996.

\bibitem{Byckling}
  E.~Byckling and K.~Kajantie,
  Nucl.\ Phys.\  B {\bf 9}, 568 (1969).


\bibitem{Kersevan}
B.~P.~Kersevan and E.~Richter-Was,
  Eur.\ Phys.\ J.\  C {\bf 39}, 439 (2005)
  [arXiv:hep-ph/0405248].

\bibitem{CUBA}
T.~Hahn,
  Comput.\ Phys.\ Commun.\  {\bf 168}, 78 (2005)
  [arXiv:hep-ph/0404043].






\end{thebibliography}
\end{document}